\documentclass{aa}  

\usepackage{graphicx}
\usepackage{txfonts}
\usepackage{xcolor}
\usepackage{orcidlink}
\usepackage{ulem}
\usepackage{hyperref} 
\usepackage{lineno}

\def\apjl {ApJL}
\def\apjs {ApJS}

\def\aap {A\&A}

\def\R200 {R_{200}}

\def\rogerii {\texttt {Roger v2.0}}

\definecolor{sele}{RGB}{255,105,180}  
\definecolor{andrea}{RGB}{77,77,255}     
\definecolor{hernan}{RGB}{255,140,0}  
\definecolor{vale}{rgb}{0.8,0,0}      
\definecolor{julian}{rgb}{0.5,0,0.5}

\begin{document} 

\authorrunning{S. Levis et al.}

  \title{Linking X-ray emission to galaxy populations in GAMA groups} 
   
  \author{Selene Levis\inst{1,2}\orcidlink{0000-0003-1887-776X},
  Hernán Muriel\inst{1,3}\orcidlink{0000-0002-7305-9500},
  Valeria Coenda\inst{1,3}\orcidlink{0000-0001-5262-3822},
  Andrea Biviano\inst{4,5}\orcidlink{0000-0002-0857-0732},
  Héctor J. Martínez\inst{1,3}\orcidlink{0000-0003-0477-5412}
  \& Martín de los Rios\inst{1}\orcidlink{0000-0003-2190-2196}}

\institute{
        Instituto de Astronomía Teórica y Experimental, CONICET - UNC, Laprida 854, X5000BGR, Córdoba, Argentina
        \and
        Facultad de Matemática, Astronomía, Física y Computación, Universidad Nacional de Córdoba, Av. Medina Allende s/n, X5000HUA, Córdoba, Argentina
        \and
        Observatorio Astronómico, Universidad Nacional de Córdoba, Laprida 854, X5000BGR, Córdoba, Argentina
        \and
        INAF-Osservatorio Astronomico di Trieste, via G.B. Tiepolo 11, 34143 Trieste, Italy
        \and
        IFPU-Institute for Fundamental Physics of the Universe, via Beirut 2, 34014 Trieste, Italy}

   \date{Received XXXX; accepted XXXX}


   \abstract
    {The evolution of galaxies is strongly influenced by their environment, but the role of the intragroup medium in shaping galaxy properties at group scales remains unclear. In particular, it is not yet established whether the presence of a hot intragroup medium produces differences in galaxy evolution beyond those associated with halo mass.}
   {We investigated how the presence of a detectable hot intragroup medium affects the colour and star-formation properties of galaxies and how these trends depend on their dynamical stage within the group.}
   {We analysed galaxies in GAMA groups with and without X-ray detection (XG and NXG, respectively) at $0.1\leq z\leq0.2$, using samples with similar halo-mass distributions. We compared them with a field control sample and classified galaxies into red, green, and blue populations using their UV–optical colour. We further classified group galaxies into dynamical classes, based on their position in projected phase space and host halo mass. We investigated galaxy populations as a function of orbital class, star-formation activity, nuclear activity, and cluster-centric distance.}
   {The galaxy population is systematically more evolved in XG than in NXG, with lower blue and star-forming fractions and higher red and passive fractions. The green fraction remains relatively stable across environments and orbital classes. However, high-mass galaxies that have recently entered the system exhibit a mild excess of green systems while retaining relatively high star-forming and low passive fractions, suggesting an intermediate stage of star-formation suppression. The active galactic nucleus fraction shows no significant environmental dependence, whereas the post-starburst fraction increases strongly in XG, reaching nearly 20\%, consistent with enhanced rapid quenching. Galaxy properties also show stronger radial segregation in XG than in NXG, with colour fractions remaining different from field values even at $R/R_{200}\sim3$, consistent with preprocessing.}
   {Despite their similar halo-mass distributions, the systematic differences between XG and NXG indicate that halo mass alone does not determine galaxy properties. The presence of a detectable hot intragroup medium is associated with enhanced environmental processing, stronger quenching, and a higher incidence of rapid transitions. These results highlight the role of the thermodynamical state of the intragroup medium, in addition to halo mass, in shaping galaxy evolution at group scales.}


\maketitle
\nolinenumbers

\section{Introduction}\label{sec:intro}

In the local Universe, galaxies are fundamentally categorised into two broad populations: red passively evolving galaxies, which are typically found in high-density regions, and blue star-forming (SF) galaxies, which predominantly inhabit lower-density environments. These two populations, often referred to as the red sequence (RS) and the blue cloud (BC), are bridged by a transition population known as green valley (GV) galaxies \citep{Wyder:2007}. These intermediate objects are essential for understanding how galaxies migrate from active star formation to quiescence. 

Observational studies have shown that GV galaxies exhibit structural, kinematic, and star-formation properties distinct from those of both SF and passive (PS) systems. These properties have been investigated over a broad range of stellar masses. In the local Universe, \citet{Belfiore:2018} analysed MaNGA galaxies in the range $9.0<\log(M_{\star}/M_{\odot})<11.0$,
and find that GV galaxies exhibit suppressed specific star formation rates (sSFRs) across their galactic extent rather than only in their central regions. Focusing on galaxies with $\log(M_{\star}/M_{\odot})\geq10.0$, \citet{Bait:2017} further report that the GV is dominated by early-type spirals (49.1\%) and lenticular (S0) galaxies (39.9\%), which highlights the close connection between the transition in star formation and morphological transformation. In the same stellar-mass range, \citet{Schawinski:2014} proposed a two-pathway quenching scenario: early-type galaxies cross the GV rapidly, with morphological transformation and quenching occurring nearly simultaneously, while late-type galaxies undergo a slower transition over several gigayears. In the local Universe ($z\leq0.2$), disc-dominated GV galaxies with $10.25<\log(M_{\star}/M_{\odot})<10.75$ frequently exhibit characteristic features such as fading discs, prominent ring structures, and loosening spiral arms \citep{Smith:2022}.

At lower masses, \citet{Sampaio:2022} studied galaxies with $9\leq\log(M_{\star}/M_{\odot})<10$ at $0.03\leq z\leq0.1$, finding that the morphological transformation of low-mass cluster galaxies can occur on relatively short timescales after crossing the virial radius. Similarly, studies of nearby S0 galaxies have identified rejuvenated SF systems down to $\log(M_{\star}/M_{\odot})\sim8.75$ (e.g. \citealt{Rathore:2022}), while \citet{Parente:2025} investigated galaxies with $\log(M_{\star}/M_{\odot})\geq9$ at $0.002<z<0.08$, finding that GV galaxies can retain substantial cold-dust emission during the quenching process.

At intermediate redshifts, the structural and star-formation properties of GV galaxies have also been characterised. For example, \citet{Mendez:2011} found that a substantial fraction of GV galaxies at $0.4<z<1.2$ already exhibit early-type morphologies, while other studies have identified suppressed central star formation in massive galaxies at $0.4<z<1.0$ \citep{Wang:2017}.\citet{Estrada-Carpenter:2023} find that massive galaxies ($\log(M_{\star}/M_{\odot})>10.2$) in the GV at higher redshifts ($0.8<z<1.65$) undergo significant structural evolution, becoming more compact through an increase in the Sérsic index and a decrease in effective radius. Thus, although the detailed spatially resolved characterisation of GV galaxies is largely restricted to the nearby Universe, the existence and evolution of the GV have been established over a considerably broader range of redshift and stellar mass.

Galaxy properties, such as stellar mass, colour, and star formation history, evolve through a complex interplay between internal secular processes, including mass quenching \citep[e.g.][]{Peng:2010}, active galactic nucleus (AGN) feedback \citep[e.g.][]{Trayford:2016, Blank:2022}, and external environmental mechanisms. In this context, galaxy groups and clusters provide natural laboratories for studying how the environment  impacts galaxy evolution.

Over the past decades, numerous studies have established strong correlations between galaxy properties and environment \citep[e.g.][]{Blanton:2005, Martinez:2006, Martinez:2008}. In particular, properties such as luminosity \citep[e.g.][]{Adami:1998, Girardi:2003, Coenda:2006}, colour \citep[e.g.][]{Balogh:2000, Baldry:2006, Martinez:2008, Venhola:2019, Levis:2025}, star formation rate \citep[e.g.][]{Muzzin:2012, Darvish:2016, Coenda:2019, Martinez:2023}, and morphology \citep[e.g.][]{Dressler:1980, Bamford:2009, Skibba:2009, Kawinwanichakij:2017, Martinez:2023} show systematic trends with cluster-centric distance. As galaxies move within these deep potential wells, several mechanisms can remove their gas reservoirs and suppress star formation. These mechanisms include ram pressure stripping \citep[e.g.][]{GG:1972, Abadi:1999, Book:2010, Steinhauser:2016, Biviano:2024}, strangulation \citep[e.g.][]{Larson:1980, Peng:2010}, and tidal interactions \citep[e.g.][]{Zwicky:1951, Gnedin:2003a, Villalobos:2014}. The properties of the hot intracluster medium (ICM) largely determine the efficiency of these processes. In rich clusters, the ICM fills a substantial fraction of the volume and produces strong X-ray emission \citep[e.g.][]{Rosati:2002}. By contrast, galaxy groups host cooler and less dense gas, making their X-ray emission intrinsically fainter. Consequently, galaxies in environments with strong X-ray emission, associated with a hot and dense ICM, may follow evolutionary pathways that differ from those in optically selected clusters.

Environmental effects are reflected in the properties and evolutionary pathways of GV galaxies. In the local Universe ($z\lesssim0.15$), environmental effects on GV galaxies have been reported down to $\log(M_{\star}/M_{\odot})\sim9$, with the fraction of GV galaxies being higher in groups and clusters than in the field at lower stellar masses \citep{Coenda:2018, Bait:2017, Sampaio:2022}. By contrast, galaxies 
more massive than $\sim10^{10.5} M_{\odot}$ are often quenched predominantly by internal 
mechanisms, such as AGN feedback, with a weaker dependence on environment 
\citep{Wright:2019, Sampaio:2023}. The evolution of GV galaxies, however, is 
not necessarily monotonic towards quiescence, as gas accretion, mergers, and feedback can lead to stalled quenching or rejuvenation \citep{Chauke:2019, Nelson:2018, Rathore:2022, Parente:2025}. 
Recent studies have linked GV evolution to satellite accretion and subsequent quenching \citep{Das:2021, Sampaio:2024, Levis:2025}.

Empirical evidence supports accelerated galaxy evolution in X-ray bright environments. For instance, \citet{Coenda:2009} showed that X-ray-selected clusters host a higher fraction of early-type galaxies at all cluster-centric distances and exhibit distinct radial trends in galaxy size, suggesting that a hot ICM accelerates quenching and structural transformation. \citet{Roberts:2016} find a lower fraction of SF and disc galaxies among satellites in X-ray bright groups and clusters than in systems with weaker X-ray emission, particularly at low stellar masses. By contrast, central galaxies appear to be insensitive to the environment. They also reported non-Gaussian velocity distributions beyond the virial radius in X-ray weak systems, indicating dynamically younger outskirts. Analysing massive quiescent galaxies at $z\leq1$, \citet{Ditrani:2025} find that galaxies in X-ray emitting groups formed earlier and over shorter timescales than those in groups undetected in X-rays or in the field. Additionally, satellites in X-ray undetected systems are systematically younger. Together, these results indicate that X-ray emission, which traces a dense hot halo, is associated with more evolved galaxy populations and more efficient environmental quenching.

Within the hierarchical structure formation paradigm, galaxy clusters grow through the continuous accretion of smaller systems and field galaxies. For example, galaxies accreted onto clusters as group members \citep[e.g.][]{McGee:2009, DeLucia:2012, Wetzel:2013, Hou:2014} are subject to different physical processes than those infalling directly from the field or along cosmic filaments \citep[e.g.][]{Colberg:1999, Ebeling:2004, Martinez:2016, Rost:2020, Kuchner:2022}. During infall, galaxies may undergo significant transformations before reaching the cluster environment, a phenomenon known as pre-processing \citep[e.g.][]{Fujita:2004, Mihos:2004}.

The outskirts of galaxy clusters have been widely studied, revealing a complex mixture of infalling SF galaxies and backsplash systems \citep[e.g.][]{Balogh:2000, Mamon:2004, Gill:2005, Rines:2006, Aguerri:2010, Mahajan:2012, Muriel:2014, Salerno:2020, Benavides:2021}. The latter correspond to galaxies that have already crossed the dense central regions of the cluster but are currently located beyond $R_{200}$, the radius within which the mean density is 200 times the critical density of the Universe. Backsplash galaxies (BSs) have experienced the effects of the environment in the inner regions of clusters during their dive-in and subsequent outward motion. However, because their interaction with the dense ICM is relatively brief compared to that of long-term cluster members, they typically exhibit characteristics intermediate between those of field galaxies and those in the cluster core \citep[e.g.][]{Muriel:2014, Ruiz:2023, Levis:2025}.

To assess the impact of environmental mechanisms on galaxy evolution, it is essential to accurately characterise the different galaxy populations inhabiting clusters and their surrounding regions. Several studies have classified galaxies based on their distribution in the projected phase-space diagram (PPSD), where every galaxy is characterised by two coordinates: the projected distance to the cluster centre in units of a given characteristic cluster size and the line-of-sight velocity relative to the cluster expressed in units of its velocity dispersion \citep[e.g.][]{Biviano:2002, Mahajan:2012, Oman:2013, Muriel:2014, Rhee:2017, Pasquali:2019}. \citet{delosRios:2021} present the code, ROGER (Reconstructing Orbits of Galaxies in Extreme Regions), which links the PPSD position of galaxies to their 3D orbital classification using machine-learning techniques.

Irrespective of the classification technique employed, misclassifications inevitably introduce contamination between PPSD classes, particularly in the central regions of clusters. \citet{Coenda:2022} show that this contamination can bias the interpretation of galaxy properties, emphasising that a clear separation between red and blue populations is required to improve observational constraints. More recently, \citet{Martinez:2025} introduced a method based on the inversion of the confusion matrix to statistically correct the distribution of galaxy properties. This approach, applicable to any PPSD-based classification, provides a robust framework for mitigating contamination.

In observational applications, ROGER has been used to investigate morphological and quenching transitions in massive SDSS X-ray clusters \citep{Martinez:2023} and in 35 OmegaWINGS clusters \citep{Muriel:2025}. However, these studies were restricted to high-mass systems, as the underlying model was trained exclusively on clusters with $M_{200} \geq 10^{15}h^{-1}\mathrm{M_{\odot}}$\footnote{$M_{200}$ denotes the mass enclosed within the radius, $R_{200}$, where the mean density is $200$ times the critical density of the Universe.}. \citet{delosRios:2026} present a new version of the ROGER code extending the method to intermediate mass systems with $M_{200} \ge 10^{13.5}$ $h^{-1}$ $\mathrm{M_{\odot}}$.

In this work, we analysed the properties of red, green, and blue galaxies in and around galaxy groups and clusters with and without X-ray emission, with the aim of improving our understanding of how the cluster environment influences galaxy evolution. We employed the new version of the ROGER code to classify galaxies in the PPSD and interpret our results following the theoretical framework presented in \citet{Coenda:2022} and \citet{Martinez:2025}. This paper is structured as follows. In Sect. \ref{sec:sample} we describe the sample and the classification methods. Section \ref{sec:results} presents our results. In Sect. \ref{sec:conclusions} we summarise our findings and discuss their implications for galaxy evolution in group environments.


\section{Observational data}\label{sec:sample}

The observational data used in this work were taken from the Galaxy And Mass Assembly (GAMA) survey\footnote{\url{https://www.gama-survey.org/}} \citep{Driver:2009,Driver:2011,Liske:2015,Baldry:2018,Driver:2022}. The GAMA project aims to use the latest generation of ground-based and space-based survey facilities to investigate cosmology and the formation and evolution of galaxies. Its core component is a spectroscopic survey covering about $286$ $\rm deg^{2}$. It is complete down to a limiting magnitude $r=19.8$ and results in a sample of $\sim 300000$ galaxies. The observations were obtained with the AAOmega multi-object spectrograph on the Anglo-Australian Telescope. The survey also incorporates and extends data from earlier spectroscopic programmes, including the Sloan Digital Sky Survey (SDSS; \citealt{York:2000}), the 2dF Galaxy Redshift Survey (2dFGRS; \citealt{Colless:2001}), and the Millennium Galaxy Catalogue (MGC; \citealt{Liske:2003}).  The galaxy samples employed in this work are taken from the GAMA Data Release 4 (GAMA DR4; \citealt{Driver:2022}).


\subsection{The group sample}

We used a sample of galaxy groups based on the catalogue presented by \citet{Popesso:2024}. Their sample was obtained from a cross-match between two datasets: the eROSITA X-ray-selected groups sample of \citet{Liu:2022}, which includes over 500 extended sources up to $z\sim1$, and the GAMA optical friends-of-friends (FoF) group catalogue \citep{Robotham:2011}, which contains approximately $7500$ groups. The cross-matched sample comprises 189 groups at $z<0.2$, each with at least five spectroscopically confirmed members and halo masses $10^{13}~{\rm M}_\odot \leq M_{200} \leq 10^{15.1}~{\rm M}_\odot$. Of these 189 systems, 32 are detected in X-rays and 157 are undetected.

In this work, we considered galaxy groups with and without X-ray emission (hereafter XG and NXG, respectively) in the redshift range $0.1 \leq z \leq 0.2$. To isolate the effects of X-ray emission from those of halo mass, we applied a Monte Carlo algorithm to select XG and NXG samples with matched $M_{200}$ distributions. 
This matching procedure limits the analysis to halos with masses in the mass range common to both, XG and NXG samples: $10^{13.2}\,\leq M_{200} \leq 10^{14.6}\,M_{\odot}$. This results in final subsamples of 25 XG and 25 NXG groups. To verify that the two $M_{200}$ distributions are consistent, we computed the p-values of the Kolmogorov–Smirnov, Anderson–Darling, and Mann–Whitney U tests. None of these tests indicates a statistically significant difference ($p<0.05$), suggesting that the mass distributions of the two samples are consistent.


\subsection{The galaxy sample}\label{subsec:galaxy_group_sample}

We selected galaxies spanning the same redshift range as our cluster samples. Photometric measurements were taken from the \textsc{LambdarPhotometry} catalogue \citep{Wright:2016}, which provides multiwavelength fluxes for each galaxy. All magnitudes were corrected for extinction and reported in the AB system. $K-$corrections to $z=0.1$ were applied following \citet{Loveday:2012}. Absolute magnitudes were computed assuming a flat cosmology with $\Omega_{\Lambda,0}=0.7$, $\Omega_{m,0}=0.3$, and $H_0=70\,{\rm km\,s^{-1}\,Mpc^{-1}}$. We used several derived galaxy properties from different GAMA-related catalogues. Stellar masses and sSFRs were taken from the \textsc{MagPhys} catalogue, while spectroscopic emission-line measurements were obtained from the \textsc{GaussFitComplex} catalogue \citep{Baldry:2018}. The sSFR values were derived from combined UV and total IR spectral energy distribution fits and represent averages over the last $0.1$ Gyr. This primary sample amounts to $9796$ objects in total. 

Since the parent sample is flux-limited and many galaxy properties correlate with luminosity, all analyses were performed using $1/V_{\rm max}$ weights \citep{Schmidt:1968} to correct for volume incompleteness, unless otherwise stated. In addition, whenever the available sample size permits, we restricted the analysis to galaxies with $M_r \leq -20$, defining a volume-complete subsample. These cases are explicitly indicated throughout the paper.

\begin{figure}[!h]
\centering
{\includegraphics[width=0.5\textwidth]{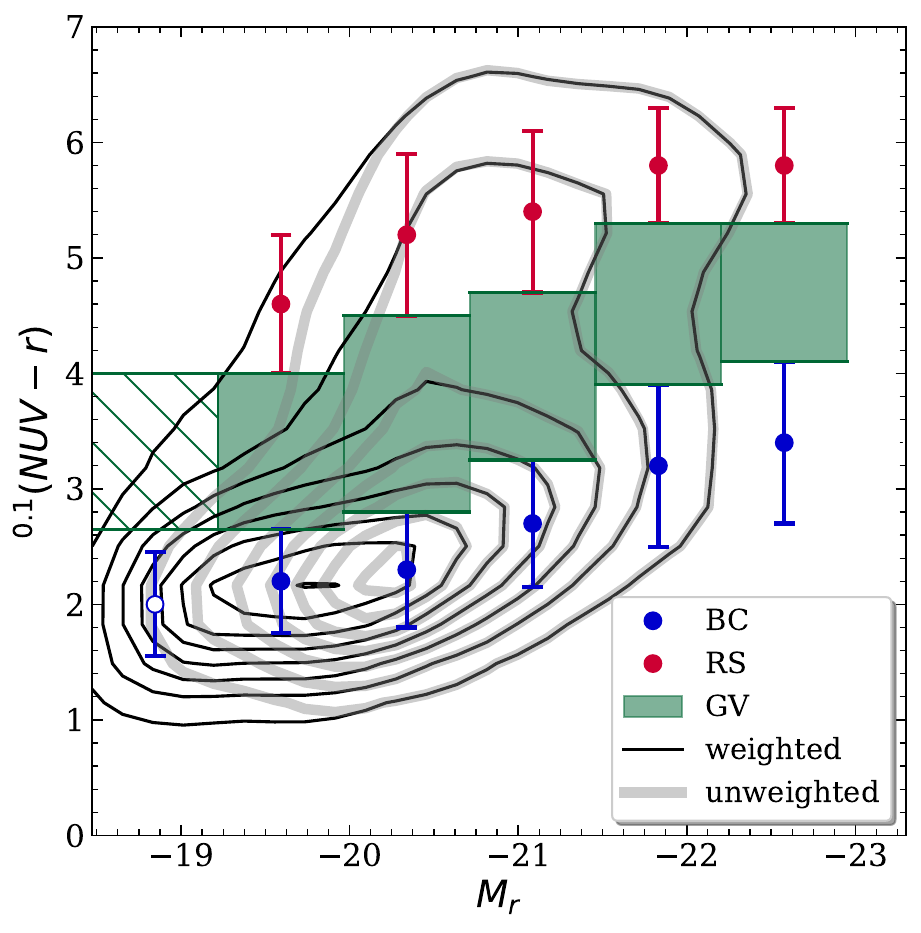}
\caption{\label{fig:gv-def} ${}^{0.1}(NUV-r)$ CMD for GAMA galaxies restricted to $0.1\leq z \leq 0.2$. The superscript 0.1 denotes the k correction to the rest-frame at z = 0.1. The solid blue and red circles indicate the centres of the Gaussian components that best fit the blue and red populations, respectively. The blue open symbol marks cases where only the BC is present. The shaded green region indicates the GV, while the hatched area marks the GZ. Isocontours of galaxy number density are shown as black lines for the weighted distribution (using the $1/V_{\rm max}$ method) and grey lines for the unweighted case.}}
\end{figure}

Figure \ref{fig:gv-def} shows the definition of the GV adopted in the UV–optical colour–magnitude diagram (CMD) for our primary galaxy sample. Ultraviolet–optical colours are commonly used to study GV galaxies, as they provide a cleaner separation between the blue SF and red PS sequences than optical colours alone \citep[e.g.][]{Salim:2014, Quilley:2022, Nyiransengiyumva:2021, Nyiransengiyumva:2026}. To delineate the GV, we followed \citet{Levis:2025}. Galaxies were divided into six absolute magnitude bins, and in each bin the distribution of $^{0.1}(NUV-r)$ colours was fitted with either a single Gaussian or the sum of two Gaussians representing the blue and red populations, using a Bayesian information criterion \citep{Schwarz:1978} to guide this decision. When two Gaussian functions provide a better description, the GV is defined as the region between one standard deviation above the mean of the blue component and one standard deviation below the mean of the red component. At low $M_{r}$, where the colour distribution is well described by a single Gaussian, we define a green zone (GZ) with the same colour limits as the GV in the nearest stellar-mass bin. Galaxies were thus classified as blue, red, or green depending on whether they lie below, above, or within the GV (or GZ) in the corresponding stellar-mass bin.

\subsection{Categorisation by environment}

From the primary sample, we first identified galaxies in and around groups (hereafter G sample) by selecting systems within projected radii $r_{\rm proj} \leq 5R_{200}$ and line-of-sight velocity offsets $|\Delta v_{\rm los}| \leq 3\sigma$ relative to the group centres. Here, $R_{200}$ and $\sigma$ denote the projected radius and line-of-sight velocity dispersion of the groups, respectively. We define the field sample (hereafter F sample) as galaxies located beyond $5R_{200}$ and with $|\Delta v_{\rm los}| > 3\sigma$ from the centres of galaxy groups with $\log(M_{200}/M_{\odot}) \geq 13$ in the redshift range $0.1 \leq z \leq 0.2$. In the flux-limited sample, there are $6873$ galaxies: the G sample contains $1008$ galaxies ($612$ in XG and $396$ in NXG), and the F sample contains $5865$ galaxies. In the volume-limited sample there are $5299$ galaxies: the G sample contains $861$ galaxies ($500$ in XG and $361$ in NXG), and the F sample contains $4438$ galaxies.

\begin{figure}[!h]
\centering
\includegraphics[width=0.49\textwidth]{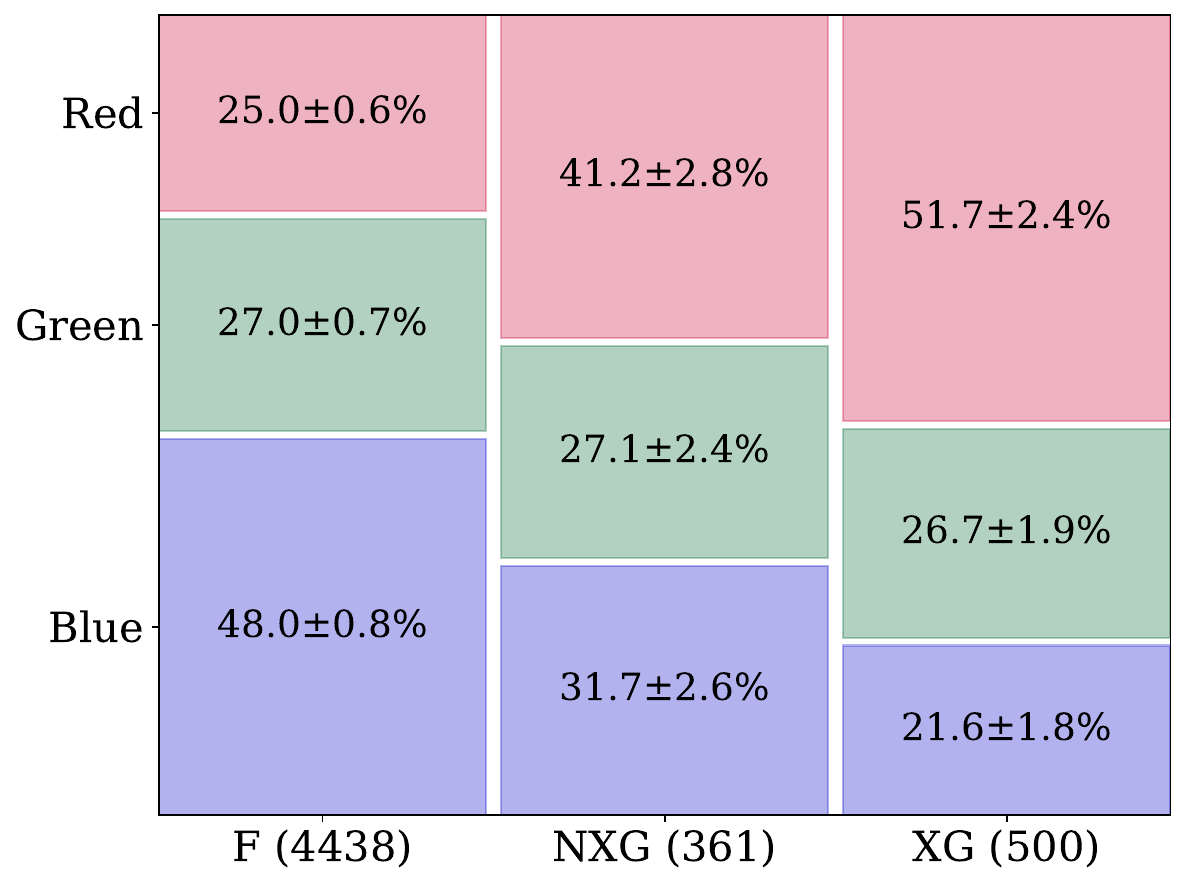}
\caption{\label{fig:perc-color}Percentage of galaxies in the blue, green, and red regions of the CMD for the different environments (F, NXG, and XG). The errors indicate bootstrap uncertainties. The number of galaxies corresponding to the volume-complete subsample in each environment is given in parentheses.}
\end{figure}

\begin{figure*}[!h]
\centering
{\includegraphics[width=0.9\textwidth]{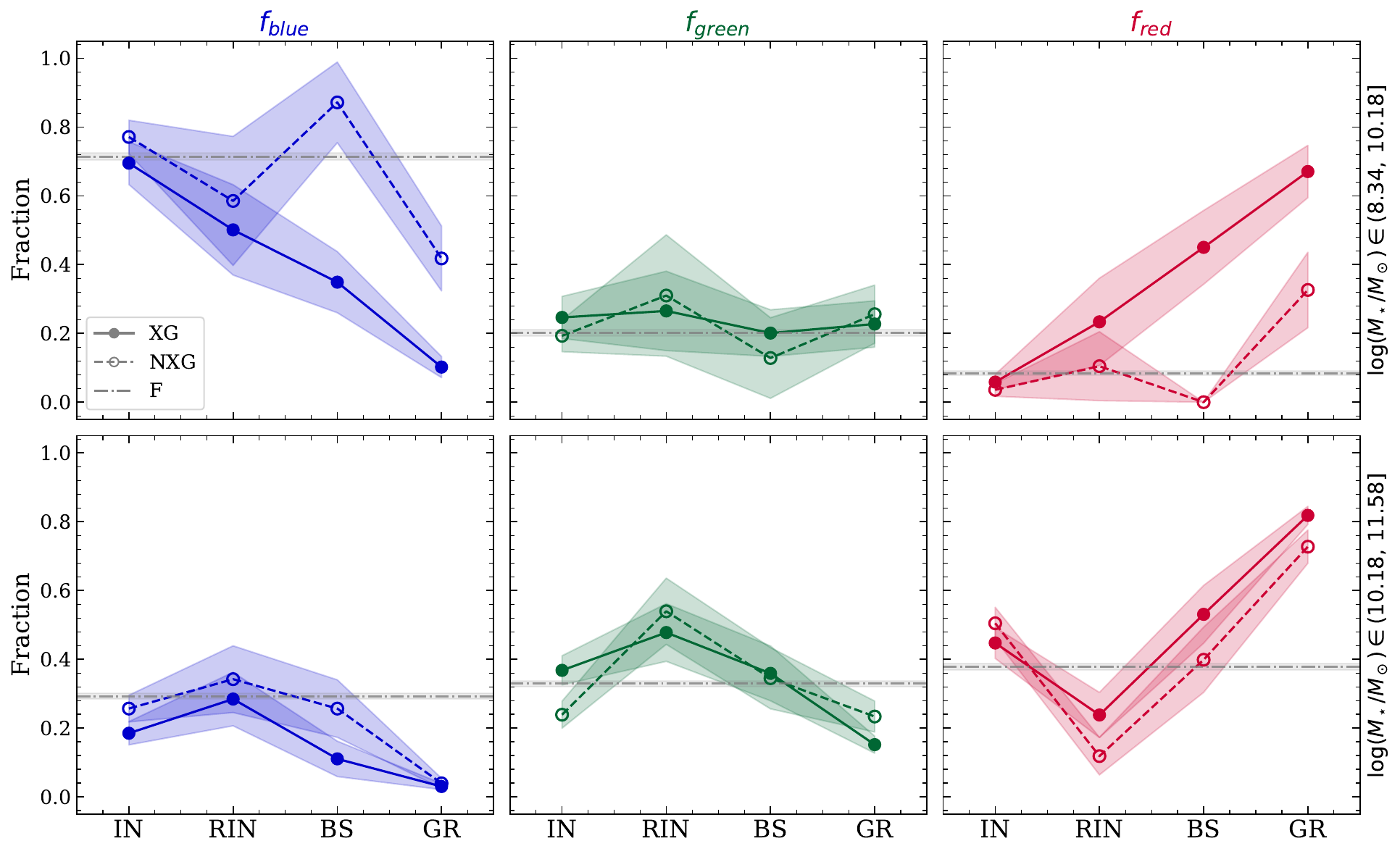}
\caption{\label{fig:types-colors} Fractions of blue ($f_{\mathrm{blue}}$, left panels), green ($f_{\mathrm{green}}$, middle panels), and red ($f_{\mathrm{red}}$, right panels) galaxies across different environments (IN, RIN, BS, GR). Results are shown for low-mass ($\log(M_*/M_\odot) \in [8.34, 10.18]$, top panels) and high-mass galaxies ($\log(M_*/M_\odot) \in [10.18, 11.58]$, bottom panels). The solid lines with filled circles indicate the X-ray-detected cluster sample (XG), while the dashed lines with open circles correspond to the non-X-ray-detected cluster sample (NXG). The horizontal dash-dotted lines represent the reference values for the field sample (F). The shaded regions indicate bootstrap uncertainties.}}
\end{figure*}

We then applied the updated \rogerii\footnote{\url{https://github.com/Martindelosrios/pyROGER}} code \citep{delosRios:2026} to classify the galaxies. We adopted a trained K-Nearest Neighbor (KNN) technique to compute the class probabilities for each galaxy. \rogerii\ estimates the probability that a galaxy belongs to any of the following orbital classes:

\begin{enumerate}
\item Group galaxies (GRs): These are galaxies that currently orbit the system as satellites and have remained in this state for more than 2 Gyr. Most of them lie within $R_{200}$, although a small fraction can be temporarily located beyond this radius due to their orbital trajectories.

\item Backsplash galaxies (BSs): These are systems that have crossed $R_{200}$ twice, once during their initial infall into the group and the second time on their way out. These galaxies are found outside
$R_{200}$.

\item Recent infallers (RINs): These are galaxies currently located within $R_{200}$ that have crossed this radius only once, during their infall, within the last 2 Gyr.

\item Infalling galaxies (INs): These are galaxies that have been outside $R_{200}$ throughout their history and are presently moving towards the system, characterised by negative radial velocities relative to the group centre.

\end{enumerate}

With the \rogerii\ computed probabilities $p_i$, $i=1,\cdots,4$, we classified the galaxies according to the following criteria. A galaxy is considered to be of the orbital class $j$, (i) if $p_j$ takes the highest value among the galaxy's four probabilities and (ii) if $p_j\geq T_j\ (M_{200})$, where $T_j\ (M_{200})$ is a threshold level which is linear on $\log(M_{200})$, as specified in \citet{delosRios:2026}. These halo-mass-dependent thresholds were computed by the authors as those that minimise the distance of the pair sensitivity-precision, $(S,P)$, to the ideal value $(1,1)$. Table \ref{tab:sample} shows the number of NXG and XG galaxies for each orbital class.

\begin{table}[ht]
\centering
\caption{Number of galaxies in the NXG and XG GAMA groups, classified according to their orbital class.}
\renewcommand{\arraystretch}{1.3}
\begin{tabular}{l c c c}
\hline
Orbital class & NXG & XG & Total \\
\hline
GR  & 118 & 239 & 357 \\
BS  &  41 &  76 & 117 \\
RIN &  35 &  66 & 101 \\
IN  & 202 & 231 & 433 \\
\hline
Total & 396 & 612 & 1008 \\
\hline
\end{tabular}
\label{tab:sample}
\end{table}


\section{Results} \label{sec:results}

Figure \ref{fig:perc-color} shows the fraction of blue, green, and red galaxies in the three environments considered: F, NXG, and XG. For this analysis, we restricted the sample to galaxies with $M_r \leq -20$, corresponding to the volume-complete subsample described in Sect. \ref{subsec:galaxy_group_sample}. The F is strongly dominated by blue galaxies (48.0\%), while the fractions of green and red systems are much smaller and comparable (27.0\% and 25.0\%, respectively). In NXG, the blue fraction decreases to 31.7\%, while the fractions of red galaxies increase to 41.2\%. The trend continues in XG, where the blue population drops to 21.6\% and the red population becomes dominant (51.7\%). The green fraction remains constant (27.1\% in NXG and 26.7\% in XG) within the uncertainties. Overall, the figure shows a clear environmental dependence of the galaxy populations, with the blue fraction decreasing and the red fraction increasing from the field to X-rayb right groups. However, the fraction of green galaxies appears independent of environment, consistent with previous studies (\citealt{Bremer:2018, Coenda:2018, Das:2021}).

\begin{figure*}[!h]
\sidecaption \includegraphics[width=12cm]{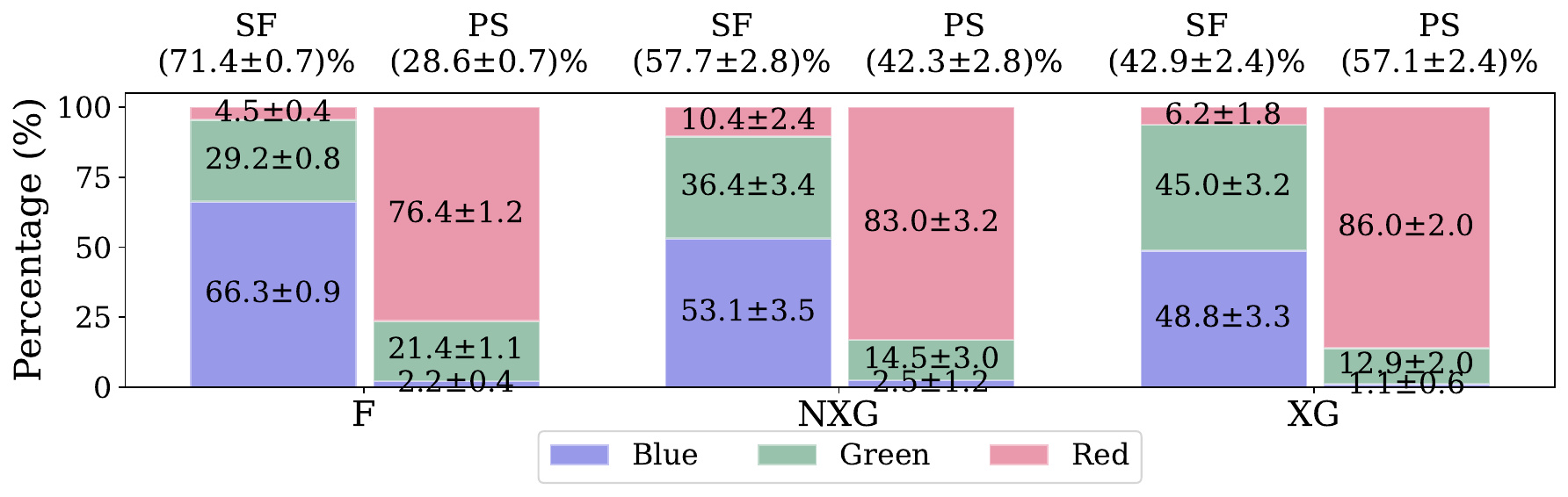}
\caption{\label{fig:ssfr-colors} Percentages of galaxies classified by colour as blue, green, or red, further separated into SF and PS populations, and according to their membership in the F, NXG, or XG samples. Uncertainties were estimated through bootstrap resampling. In the panel titles, the percentages of SF and PS galaxies are given relative to each environment.}
\end{figure*}

\begin{figure}[!h]
\centering
{\includegraphics[width=0.49\textwidth]{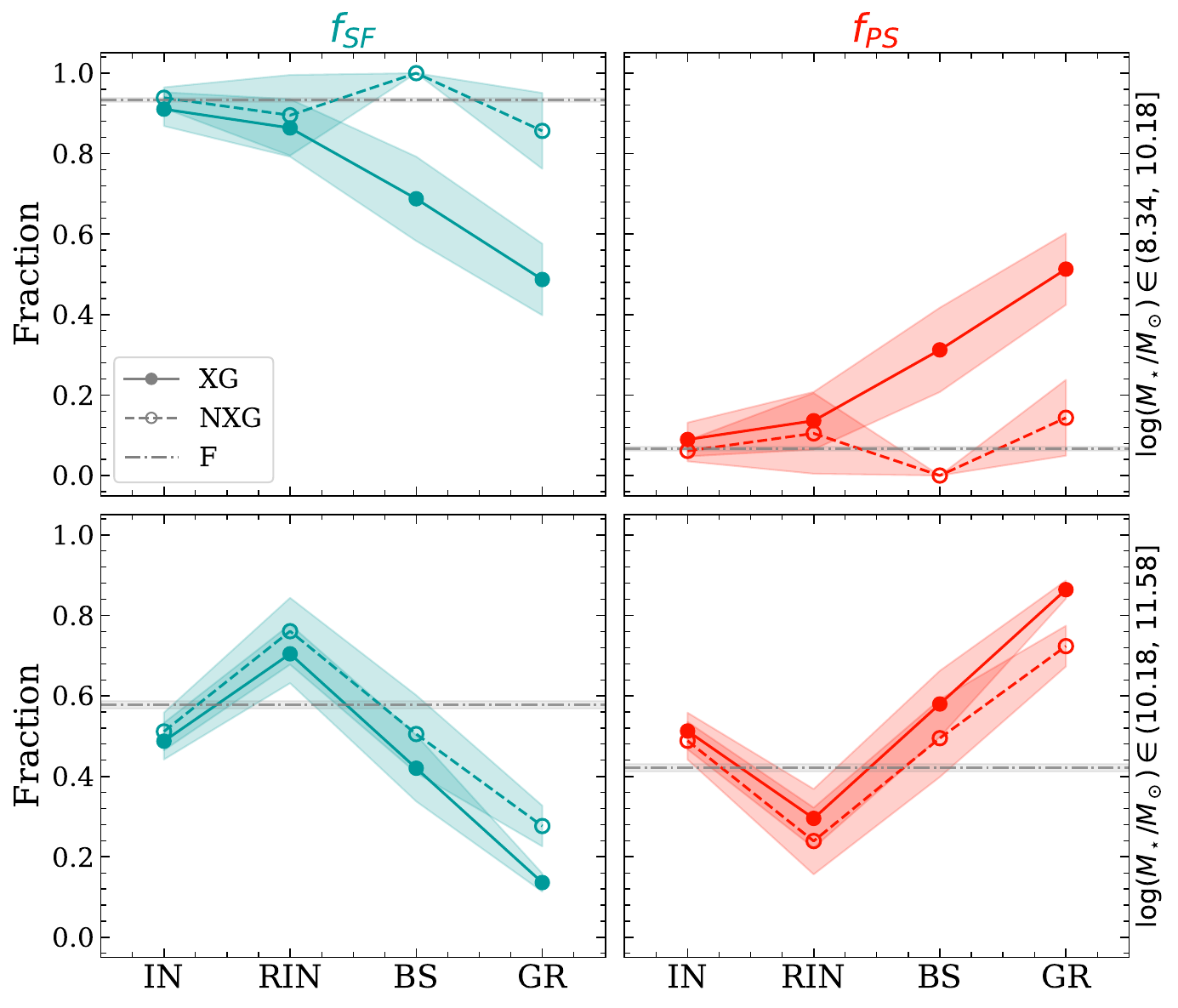}
\caption{\label{fig:types-ssfr}Fractions of SF ($f_{\mathrm{SF}}$, left panels) and PS ($f_{\mathrm{PS}}$, right panels) galaxies across different environments (IN, RIN, BS, GR). Top: Low-mass range ($\log(M_*/M_\odot) \in [8.34, 10.18]$). Bottom: High-mass range ($\log(M_*/M_\odot) \in [10.18, 11.58]$). The solid lines with filled circles denote the X-ray-detected cluster sample (XG); the dashed lines with open circles denote the non-X-ray-detected cluster sample (NXG); and the horizontal dash-dotted lines indicate the field reference values (F). The shaded regions indicate bootstrap uncertainties.}}
\end{figure}

Figure \ref{fig:types-colors} shows the fraction of blue ($f_{blue}$), green ($f_{green}$), and red galaxies ($f_{red}$) as a function of the orbital classes. For comparison, we also include the F sample. We split galaxies into two stellar mass bins: $8.34 < \log(M_*/M_\odot) \leq 10.18$ and $10.18 <\log(M_*/M_\odot) \leq 11.58$, where $\log(M_*/M_\odot)=10.18$ is the median value of the whole sample. We compute the $p$-values associated with the Kolmogorov-Smirnov (KS) test to compare the stellar-mass distributions of galaxies from NXG and XG samples. The test yields ($p>0.05$), with the KS statistics remaining below their respective critical values, indicating that the null hypothesis that the two samples are drawn from the same parent distribution cannot be rejected. To calculate each fraction, galaxies were weighted by $w=(1/V_{\rm max})\times w_{\mathrm{inv}}$, where $w_{\mathrm{inv}}$ is the inversion weight introduced by \citet{Muriel:2025} for the orbital classes. All error bars were computed by the bootstrap resampling technique. We considered only mass bins containing more than ten galaxies, with the exception of the RIN NXG point, which contains eight galaxies. Although the orbital classes considered are discrete, in Fig. \ref{fig:types-colors} we connect the points to facilitate visualisation of how the fractions vary with increasing time spent in the dense environment.

Along the orbital sequence, from recently accreted to more evolved satellite populations, the fraction of blue galaxies generally decreases, while the fraction of red galaxies increases, particularly towards the GR population. In contrast, $f_{green}$
does not exhibit a clear monotonic trend with orbital class, despite showing some variations among the different populations.

\begin{figure*}[!h]
\centering
\includegraphics[width=0.9\textwidth]{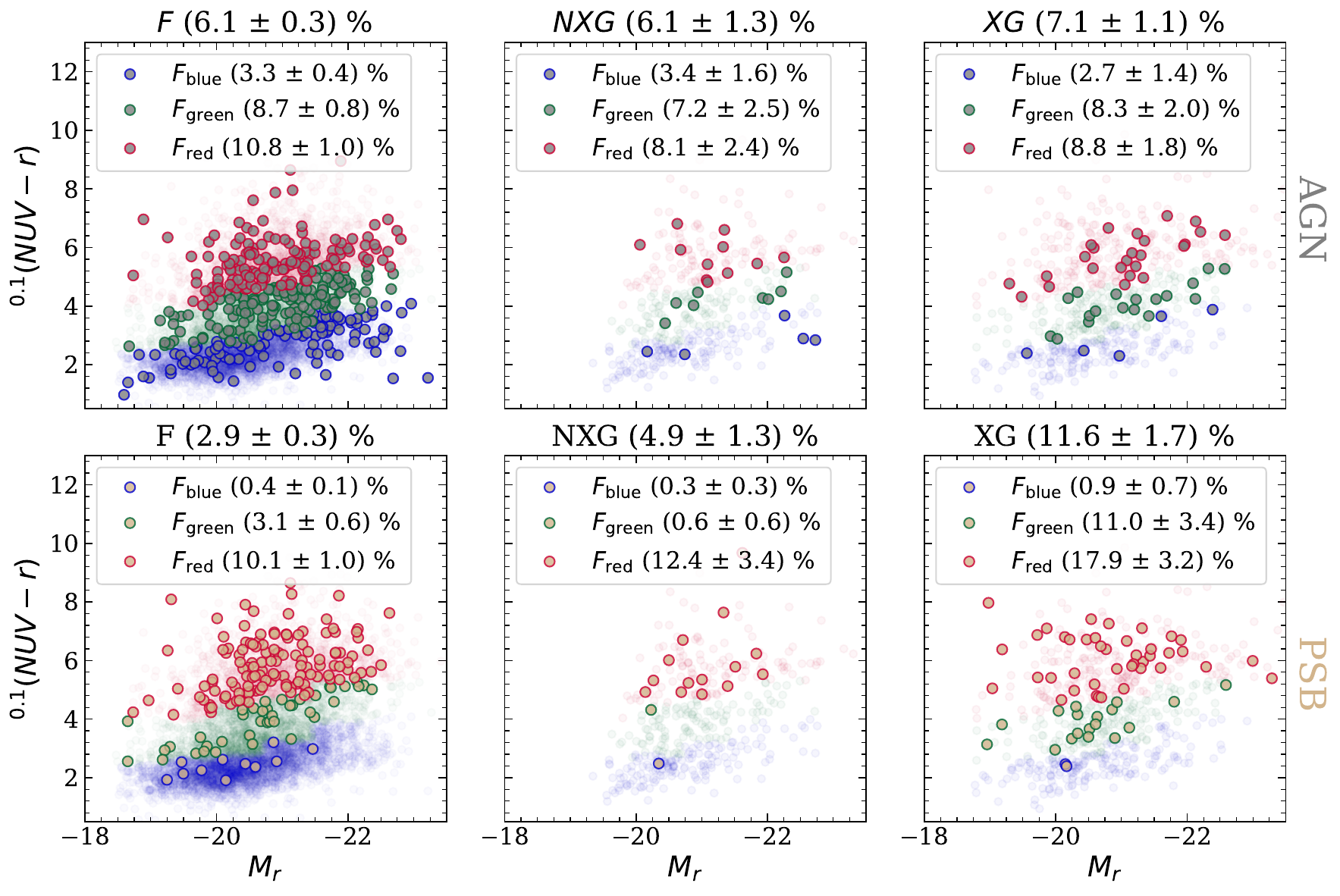}
\caption{\label{fig:agn-psb} $^{0.1}(NUV-r)$ CMDs for galaxies in different environments: F (first column), NXG (second column), and XG (third column). Semi-transparent points in the background show the full blue, green, and red galaxy populations in each environment. Top panels: AGNs populations (grey points with coloured edges) identified within the blue, green, and red regions of the CMD. Bottom panels: PSBs populations (light brown points with coloured edges) in the same regions. In each panel title, the value in parentheses indicates the fraction of AGNs or PSBs galaxies relative to the total galaxy population in that environment. The legend reports the fraction of AGNs or PSBs galaxies within the blue, green, and red subsamples of each environment.}
\end{figure*}

The colour fractions show clear dependence on both stellar mass and orbital class, with differences between XG and NXG becoming particularly evident at low stellar masses. For low-mass galaxies, the $f_{blue}$ fractions of XG and NXG are broadly comparable for IN and RIN populations. However, they diverge towards the more evolved BS and GR populations, where NXG galaxies show substantially higher blue fractions than their XG counterparts. This behaviour is accompanied by an opposite trend in $f_{red}$, with XG galaxies exhibiting higher red fractions than NXG galaxies for BS and GR. In particular, the BS population in NXG is characterised by a pronounced enhancement of $f_{blue}$ and a corresponding reduction of $f_{red}$. The $f_{green}$ fraction shows a less systematic dependence on orbital class, although an enhancement is apparent for RIN galaxies. On the other hand, both XG and NXG populations show lower $f_{blue}$ and higher $f_{red}$ than the field, with the differences becoming particularly pronounced for the more evolved BS and GR populations.

At high stellar masses, the differences between XG and NXG are generally less pronounced and do not follow a systematic trend. Nevertheless, both environments show a marked enhancement of $f_{green}$ and a corresponding deficit of $f_{red}$ among RIN galaxies, while the GR population is dominated by red galaxies and has the lowest blue fractions. 
In comparison with the F sample, the field also tends to exhibit higher $f_{blue}$ and lower $f_{red}$ fractions than XG and NXG, although the differences are generally less pronounced and vary with orbital class. Overall, these results indicate that the dependence of galaxy colour on environment is modulated by both stellar mass and orbital history, with the strongest XG-NXG differences occurring among the lower-mass, more evolved satellite populations.


\subsection{Star formation quenching}

The efficiency with which different environments drive galaxy transformation can be assessed by comparing populations with different levels of star-formation activity. We adopted a threshold of $\log(\mathrm{sSFR}/\mathrm{yr}^{-1})=-11$ to separate PS from SF galaxies. We then computed the fractions of PS and SF galaxies within the red, green, and blue populations for the three environments considered, using the volume-complete subsample described in Sect. \ref{subsec:galaxy_group_sample} These fractions are shown in Fig. \ref{fig:ssfr-colors}, where a clear environmental trend is observed: the fraction of SF galaxies decreases from the field to X-ray bright groups, while the PS fraction correspondingly increases. In the F sample, SF galaxies dominate (71.4\%), whereas in XG the population is mostly PS (57.1\%), indicating a progressive suppression of star formation in denser environments.

When separating by colour, blue galaxies are largely SF in all environments, while PS galaxies are predominantly red. From the field to group environments, the SF fraction of blue galaxies decreases slightly, whereas the PS fraction of red galaxies increases. Star-forming galaxies exhibit green fractions of $\sim 30$–$40\%$, which rise slightly from the field to group environments, and PS galaxies exhibit fractions of $\sim 10$–$20\%$, which decrease from the field to groups.

\begin{figure}[!h]
\centering
{\includegraphics[width=0.46\textwidth]{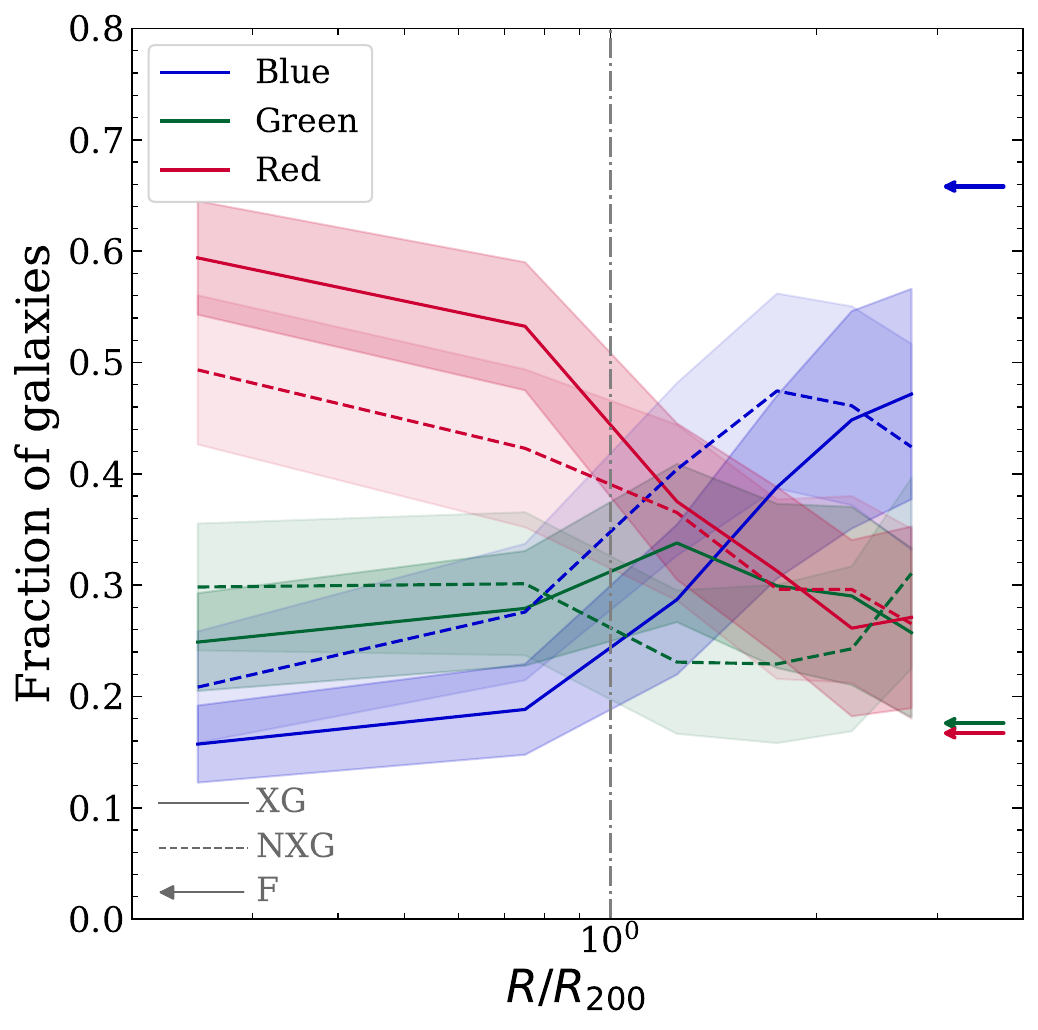}
\caption{\label{fig:segregation-color} 
Fraction of galaxies as a function of projected distance from the group centre, normalised by the group $R_{200}$, for the G sample. We classify galaxies by colour (blue, green, or red). We further divide them into XG (solid lines) and NXG (dashed lines). The shaded regions show the uncertainties estimated via bootstrap resampling. The vertical grey line marks $R_{\mathrm{proj}}/R_{200}=1$. The arrows indicate the fraction of blue, green, and red galaxies in the F sample, as in the first column of Fig.~\ref{fig:perc-color}.}}
\end{figure}

Figure \ref{fig:types-ssfr} presents the SF ($f_{\rm SF}$) and PS ($f_{\rm PS}$) fractions as a function of orbital class, with the F sample included for comparison. Galaxies are divided into the same two stellar-mass bins used in Fig. \ref{fig:types-colors}. Fractions are weighted by $w=(1/V_{\rm max})\times w_{\mathrm{inv}}$, and uncertainties were estimated through bootstrap resampling. All plotted points are based on more than ten galaxies, except for the RIN NXG point, which is based on eight galaxies.

For low-mass galaxies, $f_{\rm SF}$ is high for the IN and RIN populations and generally decreases towards the more evolved orbital classes, reaching its lowest values for GR galaxies. This decrease is particularly pronounced in XG, which also shows a corresponding increase in $f_{\rm PS}$ towards GR. In NXG, the decrease in $f_{\rm SF}$ is less pronounced, although the GR population still shows a substantially higher PS fraction than IN galaxies. Notably, the RIN population exhibits a relatively high SF fraction in both environments. At high stellar masses, the overall trends are similar, with $f_{\rm SF}$ decreasing and $f_{\rm PS}$ increasing towards GR, although the differences between XG and NXG are less systematic. The RIN population again stands out, showing a relatively high $f_{\rm SF}$ and low $f_{\rm PS}$ fraction in both environments. Overall, the figure indicates that the transition from SF to PS galaxies is more pronounced along the orbital sequence at low stellar masses, particularly in XG, while RIN galaxies retain a substantial SF component in both environments.


\subsection{Active galactic nuclei and post-starburst galaxies}

In this section, we study the occurrence of AGNs and post-starburst (PSBs) phenomena in our samples of field and GRs. Active galactic nucleus hosts were selected using the standard Baldwin, Phillips \& Terlevich (BPT) diagnostic diagram \citep{BPT:1981, Brinchmann:2004, Popesso:2006}, based on the logarithmic flux ratios $\log([\mathrm{OIII}]/\mathrm{H}\beta)$ and $\log([\mathrm{NII}]/\mathrm{H}\alpha)$, as shown in Fig.~\ref{fig:bpt-color}. We adopted the demarcation curves of \citet{Kewley:2001} and \citet{Kauffmann:2003} to separate SF galaxies, an intermediate ('composite') region, and AGNs hosts. Sources located in the intermediate region were excluded from the AGNs sample to minimise contamination. On the other hand, PSBs galaxies were identified following the spectroscopic criteria of \citet{Fritz:2014}. This classification is based on the presence or absence of $[\mathrm{OII}]$ emission and on the equivalent width (EW) of the $\mathrm{H}\delta$ absorption line. Galaxies without detectable $[\mathrm{OII}]$ emission were classified as PSBs if $\mathrm{EW}(\mathrm{H}\delta) > 3 \text{\AA}$ and as PS if $\mathrm{EW}(\mathrm{H}\delta) < 3 \text{\AA}$. Systems exhibiting $[\mathrm{OII}]$ emission were classified as emission-line galaxies. We examined the colour–magnitude distribution of our AGNs and PSBs samples in the CMD, as shown in Fig.~\ref{fig:agn-psb}. 

For AGNs, the total fraction relative to the full galaxy population is similar in all environments: $6.1\%$ in F, $6.1\%$ in NXG, and $7.1\%$ in XG. In the F sample, AGNs are most frequent in red galaxies ($10.8\%$), followed by green ($8.7\%$) and blue systems ($3.3\%$). Field AGNs span the full CMD sequence. In group environments, AGNs are slightly more frequent in green and red galaxies, with percentages of the order of $8\%$ in both NXG and XG.

Post-starburst galaxies, in contrast, show a stronger environmental dependence. Their total fraction is $2.9\%$ in F and $4.9\%$ in NXG but increases to $11.6\%$ in XG. In all environments, PSBs systems are predominantly associated with red galaxies, with percentages of $10.1\%$ in F, $12.4\%$ in NXG, and $17.9\%$ in XG, while the fraction of blue galaxies that are PSBs remains low, never exceeding $1\%$.

In the CMD, AGNs galaxies occupy the red population region with a secondary concentration in the green population, while PSBs galaxies are found mainly along the RS. This is consistent with a population undergoing recent rapid quenching. The marked increase in their total fraction in PSBs XG suggests that environments with detectable hot gas enhance the processes responsible for rapid quenching, thereby boosting the PSBs population relative to F and NXG.


\subsection{Galaxy segregation}

Several studies have shown that galaxy properties depend strongly on the local density and/or the distance from the centre of groups and clusters (e.g. \citealt{Dressler:1980, Postman:1984, Giuricin:1988, Whitmore:1993, Biviano:2002, Coenda:2006, Blanton:2007, Martinez:2008, Skibba:2009}). Here, we discuss investigating the radial segregation of galaxy properties in groups as a function of $R/R_{200}$ and explore how these trends depend on the properties of the host systems.

\begin{figure}[!h]
\centering
{\includegraphics[width=0.46\textwidth]{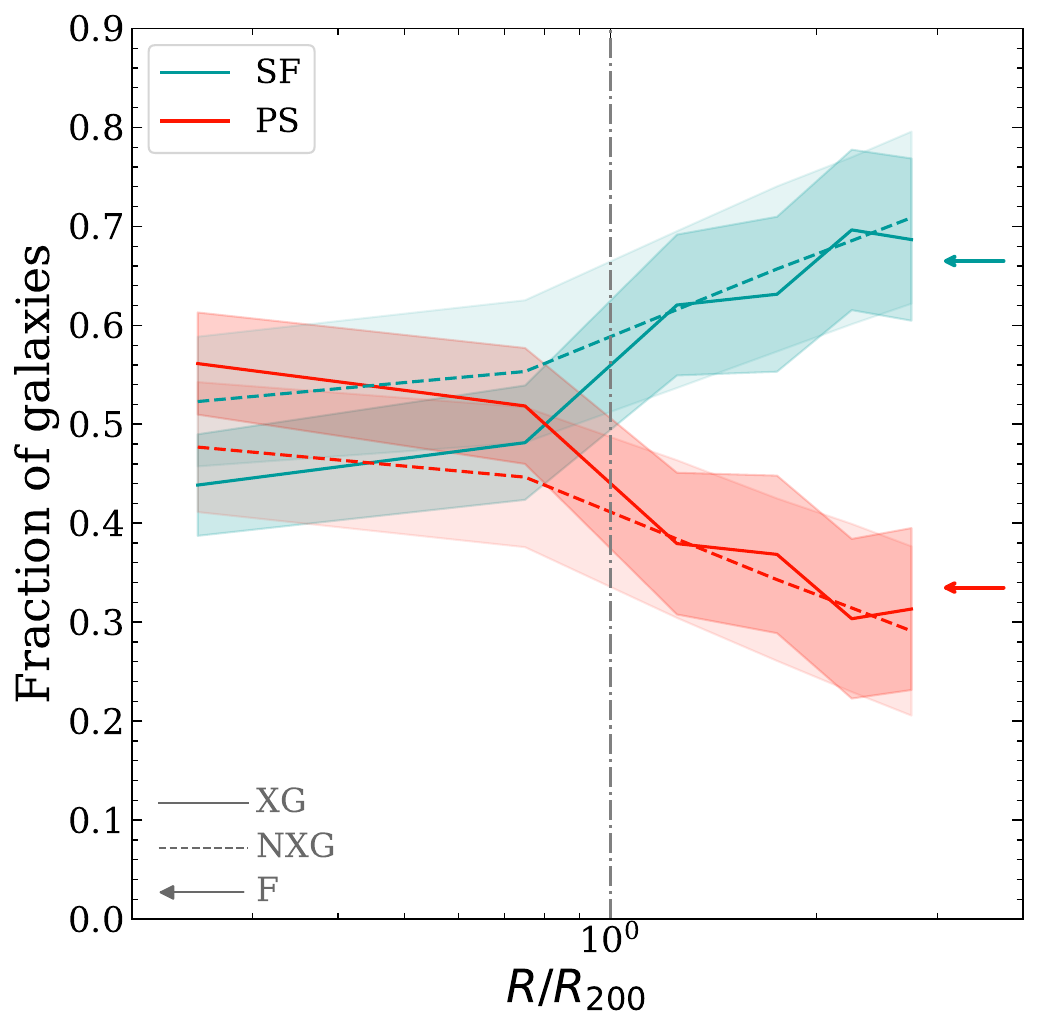}
\caption{\label{fig:segregation-sf-ps} Same as Fig. \ref{fig:segregation-color} but with galaxies classified as SF and PS and separated into the XG (solid lines) and NXG (dashed lines) groups. The shaded regions show bootstrap uncertainties. The arrows indicate the fractions in the F sample, as shown in Fig.~\ref{fig:ssfr-colors}.}}
\end{figure}

Figure \ref{fig:segregation-color} shows the fraction of blue, green, and red galaxies as a function of projected group-centric radius, for XG and NXG. Red galaxies dominate the central regions. Their fraction gradually decreases with increasing group-centric radius in both XG and NXG, though it remains consistently higher for the XG groups. Conversely, the fraction of blue galaxies increases with radius, becoming the dominant population in the outskirts. This behaviour is consistent with environmental quenching processes acting more efficiently in the dense central regions of groups, especially in systems hosting a detectable hot intragroup medium. The enhanced segregation observed in XG suggests that the presence of hot gas may play an important role in suppressing star formation in INs. The fraction of green galaxies remains intermediate and it does not show a clear dependence on radius. 
Given the uncertainties, our result of a slight increase in the fraction of green galaxies in XG groups between $R/R_{200} \sim 1$ and $R/R_{200} \sim 2$ is consistent with the findings of \citet{Levis:2026}, who, combining simulations and observations, showed that the fraction of green galaxies tends to increase towards the outer regions of galaxy systems, reaching its peak abundance beyond the virial radius. We note that, when studying colour, even out to $R/R_{200} \sim 3$ the fractions of the different populations do not reach the median values found in the field, consistent with preprocessing of galaxies in the outskirts of groups (e.g. \citealt{Mihos:2004, Fujita:2004}).

In Fig. \ref{fig:segregation-sf-ps} we show the fraction of SF and PS galaxies as a function of projected group-centric radius for XG and NXG. A clear radial trend is observed in both samples: the fraction of SF galaxies increases  with radius, while the fraction of PS galaxies decreases towards the outskirts of the systems. This behaviour closely mirrors the colour segregation shown in Fig. \ref{fig:segregation-color}, where blue galaxies become more common at larger radii and red galaxies dominate the central regions.
The gradients are again more pronounced in XG. In the inner regions ($R/R_{200}\lesssim1$), XG show a lower fraction of SF galaxies and a correspondingly higher fraction of PS systems compared to NXG. At larger radii, the fractions gradually approach the field values.
These results reinforce the interpretation that environmental mechanisms suppress star formation more efficiently in the central regions of groups, particularly in systems with a detectable hot intragroup medium.


\section{Discussion and conclusions}\label{sec:conclusions}

We analysed the properties of galaxies in GAMA groups with and without X-ray detection (XG and NXG, respectively) in the redshift range $0.1 \leq z \leq 0.2$. These systems were drawn from the sample of \citet{Popesso:2024}, who cross-matched the eROSITA extended X-ray group catalogue with the GAMA FoF groups. The group sample comprises systems with similar halo mass distributions, allowing us to study the effect of the hot intragroup medium on galaxy properties. We also compared these galaxies with a field (F) control sample drawn from GAMA.

We classified galaxies into three colour sequences: red, green, and blue, based on their UV–optical colour $^{0.1}(NUV-r)$, following the definition of \citet{Levis:2025}. We also classified group-associated galaxies in projected phase space diagram using \rogerii \citep{delosRios:2026}, which assigns each galaxy the probability of belonging to different dynamical classes: GRs, RINs, BSs, and INs. \rogerii\ uses as input the position of each galaxy in the PPSD together with the host halo mass $M_{200}$.

We find that the galaxy population becomes progressively more evolved from NXG to XG, with a decrease in the blue fraction and an increase in the red fraction. This trend is consistent with previous studies linking the presence of detectable hot gas to enhanced quenching efficiency \citep[e.g.][]{Coenda:2009, Roberts:2016, Ditrani:2025}. The green population remains significant in all environments, typically representing about $20$–$30\%$ of galaxies. When splitting the sample by orbital class, the colour fractions follow a clear sequence with increasing interaction time with the group: from IN to GR, $f_{\rm blue}$ decreases while $f_{\rm red}$ increases in both XG and NXG, whereas $f_{\rm green}$ remains relatively stable, with a mild excess of green galaxies at high stellar mass among RIN galaxies in both NXG and XG. At fixed orbital class, galaxies in XG are typically redder than their counterparts in NXG, while NXG galaxies are systematically bluer within the blue population. The persistence of a substantial green population across orbital classes is consistent with a picture in which the GV acts as a steady-state transition region, as suggested by \citet{Sampaio:2024}.

We then classified galaxies according to their star-formation activity using a threshold of $\log(\mathrm{sSFR}/{\rm yr}^{-1})$, such that galaxies above this value are classified as SF and those below it as PS. We find that the SF fraction decreases from the F sample to NXG and XG, with a corresponding increase in the PS fraction. This behaviour supports a scenario in which a denser intragroup medium enhances environmental quenching mechanisms, as suggested by \citet{Roberts:2016}.

Within the SF population, galaxies are predominantly blue, although this dominance weakens from F towards XG as the fraction of green galaxies increases slightly, while red SF galaxies remain uncommon. In contrast, the PS population is dominated by red galaxies, whose fraction increases in group environments, particularly in XG. When analysed by orbital class, the SF fraction decreases and the PS fraction increases from IN to GR, indicating progressively stronger environmental processing with increasing interaction time. At fixed orbital class, SF fractions are systematically higher in NXG than in XG, with the opposite trend for PS galaxies. For high-mass galaxies, RIN systems have the highest SF fraction of all orbital classes, accompanied by the lowest PS fraction, while their SF and PS fractions remain closer to the field values.

The combination of the colour and star-formation classifications provides further insight into the nature of the RIN population. The relatively high fraction of green galaxies among RIN systems is accompanied by a substantial SF fraction and a comparatively low PS fraction, particularly at high stellar masses. This suggests that the enhanced green fraction in RIN is not simply driven by an accumulation of already PS systems but may instead reflect galaxies undergoing an intermediate stage of star-formation suppression. Moreover, we cannot rule out that some RINs are experiencing a temporary increase in star formation. This is similar to that observed in jellyfish galaxies as a consequence of ram pressure stripping \citep{Vulcani18} and consistent with the results of \citet{Muriel:2025}, who found that RINs are the orbit class with the highest fraction of jellyfish galaxies. Towards the GR population, the decline in the green and SF fractions is accompanied by a strong increase in the red and PS fractions, suggesting a progression towards more quenched systems along the orbital sequence. Although this sequence should not be interpreted as direct evolutionary tracking of individual galaxies, the observed trends are consistent with a scenario in which environmental processing begins to affect star formation during the early stages of satellite accretion and becomes increasingly effective towards more evolved orbital populations.

We explored the environmental dependence on the nuclear activity. We find no significant environmental dependence of the AGNs fraction, which remains broadly similar in F, NXG, and XG, consistent with the weak dependence of AGNs activity on halo mass reported by \citet{Koulouridis:2024}. In contrast, the PSBs fraction exhibits a clear environmental dependence, remaining similarly low in F and NXG but increasing in XG to nearly $20\%$. Since PSBs galaxies are predominantly located around the GV and towards the RS, their enhancement in XG is consistent with recent rapid quenching. The AGNs and PSBs results suggest that the impact of the hot intragroup medium may be more closely associated with rapid quenching processes than with a general enhancement of AGNs activity.

Finally, we find a clear radial segregation of galaxy properties. Red and PS galaxies dominate the inner regions, while blue and SF galaxies become increasingly prevalent towards the outskirts and beyond $R_{200}$. These gradients are more pronounced in XG than in NXG, particularly within the virial region, indicating that the presence of a hot intragroup medium increases the efficiency of environmental processing. The colour fractions, however, do not fully reach the median values measured in the field even at $R/R_{200}\sim3$, consistent with preprocessing of galaxies in the outskirts of groups. This segregation between groups with and without X-ray emission is one of the central result of this analysis, adding a new dimension to the well-documented environmental segregation of galaxy populations from a morphological perspective \citep{Dressler:1980}, from star-formation activity \citep[e.g.][]{Balogh:1997, Balogh:1998, Lewis:2002, Biviano:2021}, and from phase-space location \citep{Oman:2013, Muriel:2014, Rhee:2017}.

We detect systematic differences between XG and NXG even though both samples share similar halo mass distributions. This indicates that halo mass alone does not fully determine galaxy properties. Instead, the presence of a detectable hot intragroup medium is associated with a more evolved galaxy population, enhanced quenching efficiency, and a higher incidence of rapid transitions. These results suggest that X-ray detectability traces not only the thermodynamical state of the intragroup medium but also the evolutionary stage of the group environment, highlighting the role of baryonic processes in shaping galaxy evolution at group scales.


\begin{acknowledgements}

We gratefully acknowledge financial support from the Argentinian institutions: Consejo Nacional de Investigaciones Científicas y Técnicas (CONICET; PIP-2022-11220210100064CO) and Secretaría de Ciencia y Tecnología de la Universidad Nacional de Córdoba (SECYT-UNC, Res. 258/53). 
This work uses data from the Galaxy And Mass Assembly (GAMA) project. GAMA is a joint European–Australasian project based around a spectroscopic campaign using the Anglo-Australian Telescope. The GAMA input catalogue is based on data taken from the Sloan Digital Sky Survey and the UKIRT Infrared Deep Sky Survey. Complementary imaging of the GAMA regions was obtained by a number of independent survey programmes, including GALEX MIS, VST KiDS, VISTA VIKING, WISE, Herschel-ATLAS, GMRT, and ASKAP, providing UV to radio coverage. GAMA is funded by the STFC (UK), the ARC (Australia), the AAO, and the participating institutions. The GAMA website is https://www.gama-survey.org/.
This project has received funding from the European Union’s HORIZON-MSCA2021-SE-01 Research and Innovation programme under the Marie Sklodowska-Curie grant agreement number 101086388 – Project acronym: LACEGAL.

\end{acknowledgements}


\bibliographystyle{aa} 
\bibliography{biblio} 

\clearpage
\onecolumn
\appendix
\section{Additional figures}

\begin{figure*}[!h]
\centering
{\includegraphics[width=1\textwidth]{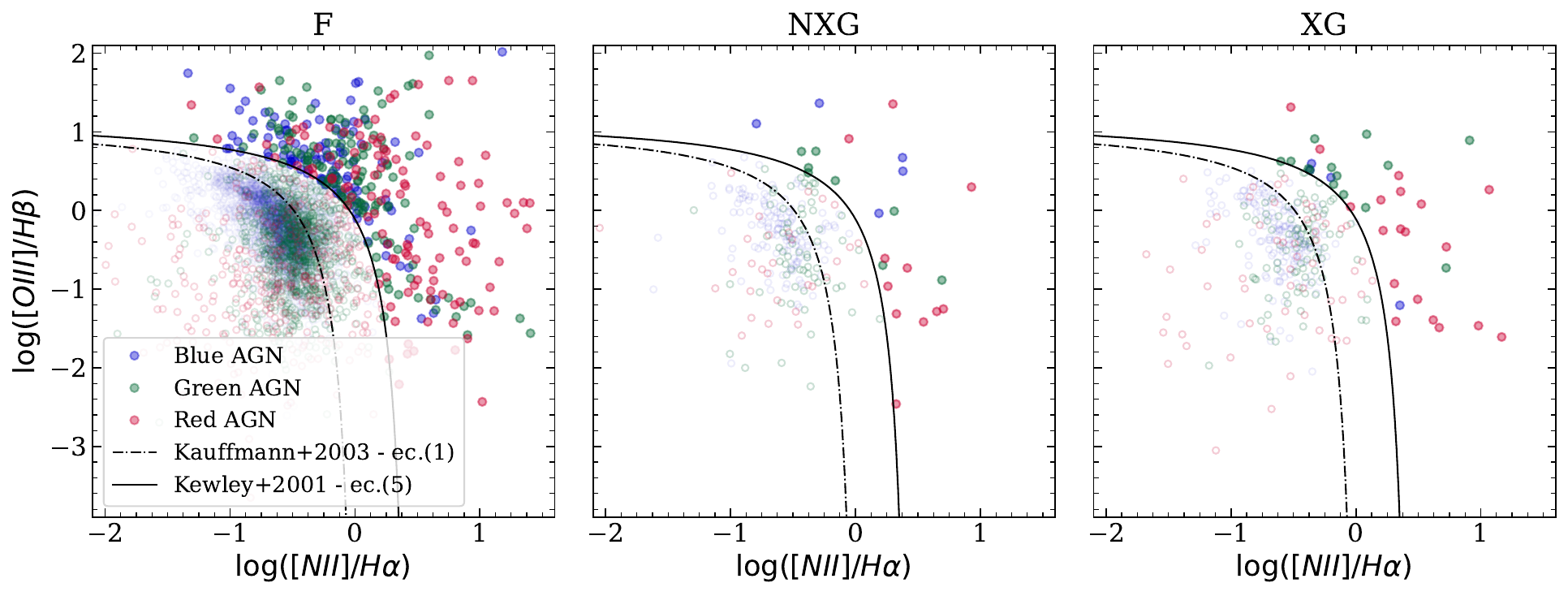}}
\caption{\label{fig:bpt-color}BPT diagnostic diagrams for galaxies in different environments: field (F), non–X-ray groups (NXG), and X-ray groups (XG). Points are colour-coded according to galaxy colour (blue, green, and red). The solid and dashed curves indicate the demarcation lines from \citet{Kauffmann:2003} and \citet{Kewley:2001}, respectively.}
\end{figure*} 


\end{document}